\documentclass[journal,hidelinks]{IEEEtran}

\usepackage{amsmath,amssymb,amsfonts}
\usepackage{graphicx}
\usepackage{booktabs}
\usepackage{array}
\usepackage{multirow}
\usepackage{url}
\usepackage{cite}
\usepackage{color}
\usepackage{balance}
\usepackage{tabularx}
\usepackage{float}
\usepackage{algorithmic}
\usepackage{orcidlink}
\usepackage{makecell}
\usepackage[caption=false,font=footnotesize]{subfig}
\graphicspath{{figures/}}
\usepackage[ruled,vlined,linesnumbered]{algorithm2e}
\usepackage{tikz}
\usetikzlibrary{arrows.meta, positioning, fit, shapes.geometric, backgrounds}

\usepackage[ruled,vlined,linesnumbered]{algorithm2e}

\usepackage{amsmath, amssymb}

\newcommand{\led}{\mathrm{LED}}
\newcommand{\R}{\mathbb{R}}

\begin{document}

\title{Latent Stability Analysis of Malware Representations Under Feature-Space Perturbations}

\author{Bamidele Ajayi\orcidlink{0000-0003-1419-9375} , Ken McGarry \orcidlink{0000-0002-9329-9835}\\ 

School of Computer Science and Engineering,\\ University of Sunderland, Sunderland, UK\\
}

\maketitle

\begin{abstract}
Static malware detectors are commonly evaluated using clean-sample metrics such as accuracy, F1, ROC AUC, and PR AUC. However, these metrics provide limited insight into how learned malware representations behave when feature vectors are perturbed, how close samples move toward uncertain decision regions, or whether compressed representations preserve security-relevant structure. This paper presents a latent-stability analysis pipeline for malware perturbation assessment in EMBER feature space. The pipeline compares full EMBER features, PCA-based compression, beta/denoising variational autoencoder representations, Mandelbrot-inspired escape-time descriptors, and a PINN-style latent-flow module. We define Latent Escape Divergence (LED) to measure changes in escape-time profiles under perturbation, and use PINNFlow-derived residual, velocity, risk, and gradient-shift metrics to characterize latent movement. Experiments are conducted on EMBER static PE feature vectors using 180,000 training samples, 180,000 test samples, and 240,000 holdout samples. Full EMBER features achieve the strongest clean classification performance with ROC AUC of 0.9962 and F1 of 0.9713, while PCA-64 is the strongest compressed baseline with ROC AUC of 0.9846 and F1 of 0.9347. The proposed VAE+Mandelbrot+PINNFlow representation does not outperform these baselines for clean classification, but it provides additional diagnostic value under controlled feature-space perturbation probes. These probes are intended for representation-stability analysis and should not be interpreted as functionality-preserving PE-level malware transformations. In particular, PINNFlow improves robustness relative to VAE+Mandelbrot across all evaluated perturbation types and yields interpretable latent-shift measurements. The results show that latent-stability diagnostics can complement conventional malware classification metrics by exposing perturbation-induced representation changes that are not captured by clean-sample performance alone.
\end{abstract}

\begin{IEEEkeywords}
Malware classification, EMBER, dimensionality reduction, variational autoencoder, complex dynamics, Mandelbrot set, physics-informed neural network, latent stability, perturbation analysis, adversarial malware.
\end{IEEEkeywords}

\section{Introduction}
Machine-learning-based malware detection research has primarily focused on engineered static features and learned representations that can classify malicious executables outside brittle signatures. Data-mining approaches were shown to detect malicious executables based on metadata, byte-derived features, and supervised machine learning early on \cite{schultz2001data,kolter2006learning}. A large public dataset and feature representation for static Windows PE malware detection was formalized by the EMBER benchmark \cite{anderson2018ember}. Temporal PE malware analysis was later supported by BODMAS \cite{yang2021bodmas}, and multi-file-format evaluation with challenge samples designed to confuse malware classifiers was introduced by EMBER2024 \cite{joyce2025ember2024}. Gradient-boosted decision trees like LightGBM \cite{ke2017lightgbm} remain competitive against more complex classifiers when trained on EMBER-style features.

Robustness to adversarial examples is critical for security-focused malware analysis, however. Adversarial machine learning studies demonstrate that even small adversarial perturbations can confuse many classifiers \cite{szegedy2014intriguing,goodfellow2015explaining,papernot2017practical}. In malware detection, these perturbations are semantically/security-critical: since instructions can often be modified in ways that preserve overall executable functionality and malicious intent, detector-facing features may change under attack, causing misclassification. Prior research has explored malware evasion with reinforcement-learning attacks \cite{biggio2013evasion}, byte insertion \cite{anderson2018learning}, black-box attacks against Windows malware detectors \cite{kolosnjaji2018adversarial}, detection-resistant PE binaries \cite{demetrio2020adversarial,demetrio2021secml}, and PE packing strategies \cite{song2020mab}. An unanswered question, however, is how malware representations change when the input is perturbed. Clean Area Under Curve (AUC) does not capture this.

PCA, kernel methods, manifold learning, and autoencoders/VAEs are common tools for compressing malware feature spaces into 2D or 3D, which can then be visualized to analyze high-dimensional behavior. Principal Component Analysis \cite{jolliffe2002pca} provides a linear dimensionality reduction baseline, kernel and manifold methods mine for nonlinear spatial structure \cite{scholkopf1998kernel,mcinnes2018umap}, and autoencoders and variational autoencoders (VAEs) learn nonlinear latent spaces with neural networks \cite{kingma2014auto,vincent2008denoising,higgins2017beta}. Still, most reduced malware representations are carefully evaluated as static inputs to classifiers. There is little work on characterizing their latent behavior as \emph{objects} under perturbation.

This paper aims to bridge that gap by building a neural-geometric framework to study latent space movement induced by input perturbations. First, it projects VAE latent coordinates into complex-coordinate pairs. Then, it computes Mandelbrot-inspired escape-time fingerprints from each pair; formally defines Latent Escape Divergence (LED); and learns a physics-informed neural network (PINN)-style latent flow field across concatenated constant-and-perturbed PE trajectories. This paper does not claim that malware ``is fractal'' or behaves according to physical law. Complex dynamics and PINNs are used as principled mathematical lenses through which to measure aspects of nonlinear stability, decision boundary complexity, and adversarial trajectory consistency.

\subsection{Contributions}
To address this gap, the proposed framework studies latent-space movement induced by malware feature perturbations. The contributions are summarized as follows.
\begin{enumerate}
    \item It formulates malware perturbation analysis as latent trajectory analysis: transformations $T_t(x)$ induce paths $z_t=f_\phi(T_t(x))$ whose stability can be measured independently of clean classification accuracy.
    \item It implements a beta/denoising VAE with KL warm-up and free-bits regularization to mitigate posterior collapse, and reports active-unit and KL diagnostics.
    \item It introduces Mandelbrot-inspired escape-time fingerprints from VAE latent coordinates and defines LED as a mathematical measure of perturbation-induced complex-dynamical instability.
    \item It develops a PINN-style latent-flow module that enforces an advection-like residual over pseudo-time perturbation trajectories and adds unique residual, velocity, risk, and gradient-shift diagnostics.
    \item It uses box-counting dimension to quantify the complexity of uncertain classifier boundary regions in reduced malware representation spaces.
    \item It provides a complete ablation on EMBER comparing Full features, PCA-64, VAE-64, VAE+Mandelbrot-64, VAE+Mandelbrot+PINNFlow-64, and PINNFlowOnly-64 under clean classification and controlled perturbation probes.
\end{enumerate}

\subsection{Positioning and Scope of the Contribution}
The results support a conservative but principled claim. Full EMBER features and PCA-64 remain stronger for clean classification. The novelty lies in a diagnostic framework for latent stability, perturbation dynamics, and boundary complexity. 

\section{Related Work and Gaps}
\subsection{Static Malware Benchmarks}
EMBER provides labeled static PE feature vectors and tooling for reproducible malware learning \cite{anderson2018ember}. BODMAS introduces temporal analysis of PE malware \cite{yang2021bodmas}, while EMBER2024 broadens benchmark evaluation to multiple file types and challenge settings \cite{joyce2025ember2024}. These datasets support standardized evaluation but are often used primarily for clean classification metrics. This paper uses EMBER as a controlled setting and introduces additional diagnostics for latent dynamics and perturbation stability.

\subsection{Malware Classifiers}
Traditional malware learning uses engineered executable features and conventional classifiers \cite{schultz2001data,kolter2006learning}. Neural malware models such as MalConv process raw bytes end-to-end \cite{raff2017malware}, but strong tree-based classifiers remain difficult to beat on tabular EMBER features. LightGBM is used as a fixed downstream classifier across representations, thereby isolating the effect of representation design rather than classifier choice.

\subsection{Adversarial Malware and Functionality Preservation}
General adversarial example research established that learned classifiers can be manipulated at test time \cite{szegedy2014intriguing,goodfellow2015explaining,papernot2017practical,biggio2013evasion}. Malware evasion differs from image perturbation because practical transformations should preserve executability and malicious intent. Anderson et al. studied reinforcement-learning evasion against static PE detectors \cite{anderson2018learning}; Kolosnjaji et al. studied adversarial malware binaries against deep learning detectors \cite{kolosnjaji2018adversarial}; Demetrio et al. surveyed and implemented practical Windows malware attacks and secml-malware \cite{demetrio2020adversarial,demetrio2021secml}; and MAB-Malware studied black-box reinforcement-learning attacks \cite{song2020mab}.The default perturbations are feature-space probes, not functionality-preserving binary transformations. The framework is designed to accept external clean/perturbed feature pairs generated by raw-PE workflows, but such external PE-level evaluation is left for future work.

\subsection{Latent Representation Learning}
PCA remains a strong baseline for dimensionality reduction \cite{jolliffe2002pca}; kernel PCA and UMAP capture nonlinear structure \cite{scholkopf1998kernel,mcinnes2018umap}. VAEs provide probabilistic latent representations \cite{kingma2014auto}, denoising autoencoders encourage robustness to corrupted inputs \cite{vincent2008denoising}, and beta-VAE regularizes latent factors \cite{higgins2017beta}. The gap addressed here is that malware latent representations are often evaluated by downstream AUC/F1 without measuring the dynamics or stability of the latent space itself.

\subsection{Complex Dynamics, Fractal Geometry, and PINNs}
The Mandelbrot set is a canonical object in complex dynamics, where escape-time iteration reveals nonlinear stability and boundary complexity \cite{mandelbrot1982fractal,devaney1989chaotic}. Box-counting dimension estimates geometric complexity of sets and boundaries \cite{falconer1990fractal}. PINNs impose residual constraints during neural learning \cite{raissi2019pinn}, but can suffer from optimization and conditioning challenges \cite{krishnapriyan2021failure}. This paper adapts these tools cautiously: Mandelbrot-style escape-time is used as a nonlinear sensitivity descriptor of latent coordinates, and PINNFlow is used as a dynamics-informed regularizer rather than a claim of physical law.

\begin{table*}[t]
\centering
\caption{Prior work, open gap, and contribution of this paper.}
\label{tab:gaps}
\scriptsize
\begin{tabularx}{\textwidth}{p{0.27\textwidth} p{0.34\textwidth} X}
\toprule
\textbf{Prior work} & \textbf{Open gap} & \textbf{Contribution in this paper} \\
\midrule
EMBER, BODMAS, and EMBER2024 provide malware benchmarks \cite{anderson2018ember,yang2021bodmas,joyce2025ember2024}. & Evaluation is often dominated by clean detection metrics such as F1 and AUC. & Adds latent-stability, escape-divergence, PINN-flow, and boundary-complexity diagnostics. \\
LightGBM and raw-byte neural models provide strong malware classifiers \cite{ke2017lightgbm,raff2017malware}. & Strong classifiers do not directly explain how compressed representations move under perturbation. & Holds the downstream classifier fixed while comparing Full, PCA, VAE, complex-dynamical, and PINNFlow representations. \\
Adversarial malware work studies functionality-preserving evasion \cite{anderson2018learning,kolosnjaji2018adversarial,demetrio2020adversarial,demetrio2021secml,song2020mab}. & Attack success is often measured without characterizing latent dynamical signatures. & Defines LED and PINNFlow residual/velocity/risk-shift metrics for perturbation-induced instability. \\
PCA, kernel PCA, UMAP, autoencoders, and VAEs reduce dimensionality \cite{jolliffe2002pca,scholkopf1998kernel,mcinnes2018umap,kingma2014auto}. & Reduced representations are usually evaluated by classifier performance, not geometry or dynamics. & Introduces Mandelbrot-inspired escape-time fingerprints and box-counting boundary diagnostics. \\
PINNs enforce residual constraints for physical systems \cite{raissi2019pinn,krishnapriyan2021failure}. & Malware has no physical governing equation, so literal PINN claims are inappropriate. & Uses a PINN-style residual only as a dynamics-informed prior over pseudo-time perturbation trajectories. \\
\bottomrule
\end{tabularx}
\end{table*}

\section{Principled Framework}
\subsection{Why Latent Dynamics Matter}
Let $x\in\R^d$ be an EMBER feature vector and let $T_t$ be a perturbation operator indexed by pseudo-time or intensity $t\in[0,1]$, with $T_0(x)=x$ and $T_1(x)=x'$. A representation map $f_\phi:\R^d\rightarrow\R^m$ induces a latent trajectory
\begin{equation}
    z_t = f_\phi(T_t(x)).
\end{equation}
For a security model, the key object is not only the point $z_0$ but the path $\{z_t\}_{t=0}^1$: evasion occurs when perturbation moves a sample toward or across an unstable decision region while preserving security semantics. Clean AUC measures separability at $t=0$; latent dynamics measure the response of the representation and decision function along a perturbation path.

This view motivates three diagnostics. First, displacement and neighborhood overlap quantify geometric stability. Second, LED measures how nonlinear complex-dynamical stability changes under perturbation. Third, PINNFlow residuals measure whether perturbation movement is consistent with a learned latent transport structure. Together, these diagnostics allow the study of representation stability even when clean classification does not improve.

\subsection{VAE with Anti-Collapse Regularization}
Let a beta/denoising VAE encode $x$ to $q_\phi(z|x)=\mathcal{N}(\mu_\phi(x),\operatorname{diag}(\sigma_\phi^2(x)))$. The reparameterized latent is
\begin{equation}
    z = \mu_\phi(x)+\sigma_\phi(x)\odot\epsilon, \quad \epsilon\sim\mathcal{N}(0,I).
\end{equation}
The basic objective is
\begin{equation}
    \mathcal{L}_{\mathrm{VAE}}=\mathcal{L}_{\mathrm{rec}}+\beta_t\,\widetilde{D}_{\mathrm{KL}}(q_\phi(z|x)\Vert p(z)).
\end{equation}
To address KL collapse, $\beta_t$ is warmed up linearly from $0$ to $\beta=0.5$ over 15 epochs, and a free-bits floor of 0.01 is applied to per-dimension KL terms. Denoising is used by corrupting the input while reconstructing the clean target \cite{vincent2008denoising}. Unlike the previous beta=2.0 configuration, the run uses a 64-dimensional latent space, KL warm-up, and free bits.

\subsection{Mandelbrot-Inspired Escape-Time Descriptor}
The VAE latent vector $z\in\R^m$ is robustly standardized using training-set medians and interquartile ranges. Adjacent coordinates are mapped into complex values
\begin{equation}
    c_j = \rho\tanh(\tilde{z}_{2j-1}) + i\rho\tanh(\tilde{z}_{2j}),
\end{equation}
where $\rho$ bounds the coordinate scale. For each $c_j$, the escape-time iteration is
\begin{equation}
    w_{n+1}^{(j)}=(w_n^{(j)})^2+c_j, \quad w_0^{(j)}=0.
\end{equation}
The normalized escape time is
\begin{equation}
    \tau_j(z)=\frac{1}{N}\min\{n: |w_n^{(j)}|>r\},
\end{equation}
with $N=64$ and $r=2.0$; coordinates that do not escape are assigned $\tau_j=1$. The vector $\tau(z)$ and its summary statistics form a complex-dynamical stability fingerprint. The interpretation is precise but limited: escape time captures nonlinear sensitivity of the latent coordinate under an iterative stability test, not malware semantics by itself.

\subsection{Latent Escape Divergence}
For a clean sample $x$ and perturbed sample $x'=T_1(x)$, LED is defined as
\begin{equation}
    \led(x,x') = \frac{1}{K}\left\Vert \tau(f_\phi(x'))-\tau(f_\phi(x))\right\Vert_1.
\end{equation}
LED is therefore the mean absolute change in normalized escape-time profile. It measures perturbation-induced change in complex-dynamical latent stability. A high LED means the perturbation changed the sample's escape-time fingerprint even if the classifier score or Euclidean distance does not fully reflect that change.

\subsection{PINNFlow as Structured Latent Transport}
PINNFlow learns a latent velocity field $v_\theta(z,t)$ and scalar risk potential $u_\theta(z,t)$ over pseudo-time perturbation trajectories. Given latent states $z_0=f_\phi(x)$ and $z_1=f_\phi(x')$, a pseudo-trajectory is
\begin{equation}
    z_t=(1-t)z_0+t z_1.
\end{equation}
Inspired by PINN residual learning \cite{raissi2019pinn}, PINNFlow uses an advection-style residual
\begin{equation}
    R_\theta(z_t,t)=\frac{\partial u_\theta}{\partial t}+v_\theta(z_t,t)^\top \nabla_z u_\theta(z_t,t).
\end{equation}
The training objective is
\begin{equation}
\mathcal{L}_{\mathrm{flow}}=\lambda_T\mathcal{L}_{\mathrm{transition}}+\lambda_y\mathcal{L}_{\mathrm{sup}}+\lambda_R\|R_\theta\|_2^2+\lambda_E\|v_\theta\|_2^2.
\end{equation}
This enforces structure by discouraging arbitrary risk-potential changes along a learned flow field. In the run, $\lambda_T=1.0$, $\lambda_y=1.0$, $\lambda_R=0.5$, and $\lambda_E=10^{-4}$. PINNFlow contributes unique diagnostics: residual shift, velocity-norm shift, risk-probability shift, gradient-norm shift, and predicted-displacement shift.

\subsection{Boundary Complexity via Box Counting}
Let $s(z)$ be a classifier score. The uncertain boundary set is approximated by
\begin{equation}
    \mathcal{B}_\alpha=\{z: |s(z)-0.5|\leq q_\alpha\},
\end{equation}
where $q_\alpha$ keeps the most uncertain fraction of samples. The box-counting dimension is estimated from
\begin{equation}
    D_B \approx \frac{\Delta \log N(\epsilon)}{\Delta \log(1/\epsilon)},
\end{equation}
where $N(\epsilon)$ is the number of occupied boxes of side length $\epsilon$. Fractal descriptors are relevant because evasion often exploits irregular or high-curvature decision regions: more intricate uncertain boundaries can imply greater sensitivity to small representation movements. The paper uses this as a descriptive statistic, not as proof of fractal malware semantics.

\section{Experimental Design}
\subsection{Dataset and Split}
\begin{figure}[t]
\centering
\includegraphics[width=\columnwidth]{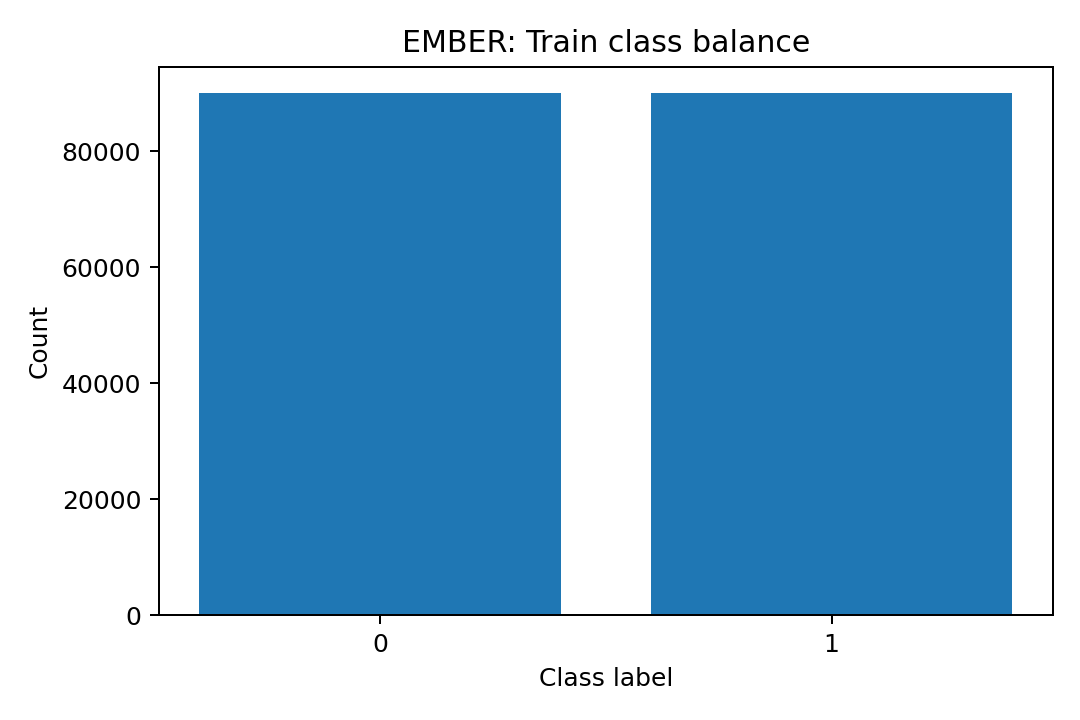}
\caption{EMBER training class balance.}
\label{fig:classbalance}
\end{figure}

The experiment uses EMBER-style static PE feature vectors with $d=2381$. The dataset split contains 180,000 training samples, 180,000 test samples, and 240,000 holdout samples. Fig.~\ref{fig:classbalance} shows balanced train labels. The configuration uses latent dimension 64, beta=0.5, KL warm-up over 15 epochs, free bits 0.01, denoising noise standard deviation 0.03, 40 VAE epochs, 25 PINNFlow epochs, and LightGBM downstream classification.

\begin{algorithm}
\caption{Latent-Stability Malware Representation and Evaluation Pipeline}
\label{alg:latent_stability_pipeline}
\small
\DontPrintSemicolon
\SetKwInput{KwRequire}{Require}
\SetKwInput{KwEnsure}{Ensure}

\KwRequire{
Feature dataset $\mathcal{D}=\{(x_i,y_i)\}_{i=1}^{n}$, latent dimension $k$,
perturbation set $\mathcal{T}$, classifier family $\mathcal{C}$.
}
\KwEnsure{
Clean-test metrics, perturbation-stability metrics, robustness summaries,
and saved experimental artifacts.
}

\BlankLine
// \textbf{Phase 1: Data Preparation}\;
Load EMBER/BODMAS-style feature vectors and labels\;
Remove invalid or unlabeled samples where applicable\;
Split data into training, test, and holdout sets:
$\mathcal{D}_{tr},\mathcal{D}_{te},\mathcal{D}_{ho}$\;
Fit scaler on $\mathcal{D}_{tr}$ and transform all splits\;

\BlankLine
// \textbf{Phase 2: Representation Learning}\;
Set $R_{\mathrm{Full}} \leftarrow X$\;
Compute $R_{\mathrm{PCA}} \leftarrow \mathrm{PCA}_{k}(X)$\;
Train beta/denoising VAE with KL warm-up and free-bits regularization\;
Encode samples using the VAE latent mean:
$Z \leftarrow \mu_{\phi}(X)$\;
Compute escape-time profile $\tau(Z)$ from bounded complex latent pairs\;
Construct $R_{\mathrm{Esc}} \leftarrow [Z,\tau(Z),s(\tau(Z))]$\;

\BlankLine
// \textbf{Phase 3: PINNFlow Feature Construction}\;
Generate controlled perturbation pairs $(x,T(x))$ for $T\in\mathcal{T}$\;
Encode pairs as latent transitions $(z,z')$\;
Train PINNFlow using transition, supervised, residual, and energy losses\;
Extract flow metrics:
velocity, risk probability, residual, gradient, and displacement\;
Construct $R_{\mathrm{PINN}} \leftarrow [R_{\mathrm{Esc}},\mathrm{FlowMetrics}(Z)]$\;

\BlankLine
// \textbf{Phase 4: Clean Classification Ablation}\;
Set representation list
$\mathcal{R}\leftarrow
\{R_{\mathrm{Full}},R_{\mathrm{PCA}},Z,R_{\mathrm{Esc}},R_{\mathrm{PINN}}\}$\;

\ForEach{$R \in \mathcal{R}$}{
    Train classifier $h_R\in\mathcal{C}$ on $R_{tr}$ and $y_{tr}$\;
    Predict labels and scores on $R_{te}$\;
    Compute Accuracy, Precision, Recall, F1, ROC AUC, and PR AUC\;
    Estimate confidence intervals using bootstrap resampling\;
}
Apply paired McNemar tests between selected classifiers\;

\BlankLine
// \textbf{Phase 5: Perturbation-Stability Evaluation}\;
\ForEach{$T \in \mathcal{T}$}{
    Generate perturbed holdout features $X'_{ho}\leftarrow T(X_{ho})$\;
    Encode $X_{ho}$ and $X'_{ho}$ into each learned representation\;
    Compute L2 displacement, cosine similarity, and kNN overlap\;
    Compute $\mathrm{LED}(x,x')=\frac{1}{K}\|\tau(z')-\tau(z)\|_{1}$\;
    Compute PINNFlow residual, velocity, risk, and gradient shifts\;
    Evaluate classifier robustness on perturbed representations\;
}

\BlankLine
// \textbf{Phase 6: External PE-Pair Evaluation}\;
\If{external clean/perturbed feature pairs are provided}{
    Load $(X_{\mathrm{clean}},X_{\mathrm{perturbed}},y)$ from NPZ file\;
    Apply the fitted scaler and representation pipeline\;
    Compute LED, PINNFlow shifts, and perturbed-classification metrics\;
}

\BlankLine
// \textbf{Phase 7: Artifact Generation}\;
Save configuration, trained models, VAE diagnostics, classification tables,
stability summaries, robustness results, and figures\;

\Return experimental metrics and saved artifacts\;

\end{algorithm}

\begin{table}[t]
\caption{Key Experimental Hyperparameters}
\centering
\begin{tabular}{ll}
\hline
Parameter & Value \\
\hline
Latent dimension & 64 \\
VAE beta & 0.5 \\
KL warm-up & 15 epochs \\
Free bits & 0.01 \\
Denoising noise std. & 0.03 \\
VAE epochs & 40 \\
PINNFlow epochs & 25 \\
PINN $\lambda_T,\lambda_y,\lambda_R,\lambda_E$ & 1.0, 1.0, 0.5, $10^{-4}$ \\
Classifier & LightGBM \\
\hline
\end{tabular}
\end{table}

\subsection{Representations}
Six representations are evaluated:
\begin{enumerate}
    \item \textbf{Full}: all 2381 EMBER features.
    \item \textbf{PCA-64}: 64-dimensional PCA representation.
    \item \textbf{VAE-64}: deterministic VAE latent mean.
    \item \textbf{VAE+Mandelbrot-64}: VAE latent vector plus escape-time profile and summary statistics.
    \item \textbf{VAE+Mandelbrot+PINNFlow-64}: VAE+Mandelbrot features plus PINNFlow diagnostics.
    \item \textbf{PINNFlowOnly-64}: only the six PINNFlow diagnostic features.
\end{enumerate}

\subsection{Perturbation Probes}
Five controlled feature-space probes are evaluated on holdout samples: Gaussian noise, sparse dropout, sparse injection, benign-centroid movement, and an EMBER byte-histogram/entropy-histogram smoothing proxy. These are diagnostic probes only. They should not be described as functionality-preserving PE attacks. A ready future path is to generate raw PE clean/perturbed pairs externally, re-extract EMBER features, and feed them into the same evaluation interface.

\begin{figure*}[!t]
\centering
\resizebox{0.96\textwidth}{!}{
\begin{tikzpicture}[
    >=Latex,
    node distance=1.35cm and 1.75cm,
    block/.style={
        rectangle,
        rounded corners,
        draw=black,
        thick,
        align=center,
        text width=3.0cm,
        minimum height=0.95cm,
        font=\footnotesize,
        fill=white
    },
    wideblock/.style={
        rectangle,
        rounded corners,
        draw=black,
        thick,
        align=center,
        text width=4.1cm,
        minimum height=1.0cm,
        font=\footnotesize,
        fill=white
    },
    smallblock/.style={
        rectangle,
        rounded corners,
        draw=black,
        thick,
        align=center,
        text width=2.8cm,
        minimum height=0.9cm,
        font=\footnotesize,
        fill=white
    },
    output/.style={
        trapezium,
        trapezium left angle=70,
        trapezium right angle=110,
        draw=black,
        thick,
        align=center,
        text width=3.1cm,
        minimum height=0.9cm,
        font=\footnotesize,
        fill=white
    },
    arrow/.style={->, thick}
]

\node[block] (data) {Input Feature Data\\EMBER / BODMAS\\Static PE Vectors};

\node[block, right=of data] (split) {Train / Test / Holdout\\Split};

\node[block, right=of split] (scale) {Feature Cleaning\\and Scaling};

\node[smallblock, below left=1.6cm and 1.2cm of scale] (full) {Full Feature\\Representation};

\node[smallblock, below=1.6cm of scale] (pca) {PCA\\$d \rightarrow k$};

\node[smallblock, below right=1.6cm and 1.2cm of scale] (vae) {Beta/Denoising VAE\\$z=\mu_{\phi}(x)$};

\node[smallblock, below=1.25cm of vae] (escape) {Escape-Time\\Profile\\$\tau(z)$};

\node[smallblock, below=1.25cm of escape] (pinn) {PINNFlow Metrics\\Residual, Velocity,\\Risk, Gradient};

\node[wideblock, below=2.75cm of pca] (repr) {Representation Set\\
Full, PCA, VAE,\\
VAE+Escape,\\
VAE+Escape+PINNFlow};

\node[block, right=4.0cm of repr] (clf) {Downstream\\Classifier\\LightGBM / HGB / RF / LR};

\node[block, right=of clf] (clean) {Clean-Test\\Evaluation\\Accuracy, Precision,\\Recall, F1, ROC AUC, PR AUC};

\node[block, below=1.25cm of clean] (stats) {Statistical\\Analysis\\Bootstrap CIs\\McNemar Tests};

\node[output, below=1.25cm of stats] (outputs) {Saved Outputs\\CSV, JSON, Figures,\\Model Artifacts};

\draw[arrow] (data) -- (split);
\draw[arrow] (split) -- (scale);

\draw[arrow] (scale) |- (full);
\draw[arrow] (scale) -- (pca);
\draw[arrow] (scale) |- (vae);

\draw[arrow] (vae) -- (escape);
\draw[arrow] (escape) -- (pinn);

\draw[arrow] (full) |- (repr);
\draw[arrow] (pca) -- (repr);
\draw[arrow] (vae) |- (repr);
\draw[arrow] (escape.west) -| (repr.east);
\draw[arrow] (pinn.west) -| (repr.east);

\draw[arrow] (repr) -- (clf);
\draw[arrow] (clf) -- (clean);
\draw[arrow] (clean) -- (stats);
\draw[arrow] (stats) -- (outputs);

\node[
    draw=black,
    dashed,
    rounded corners,
    fit=(full)(pca)(vae)(escape)(pinn)(repr),
    inner sep=0.35cm,
    label={[font=\footnotesize]above:Latent Representation Construction}
] {};

\end{tikzpicture}
}
\caption{Overall architecture of the malware latent-stability analysis pipeline. Static PE feature vectors are cleaned and scaled, then passed through full-feature, PCA, VAE, escape-time, and PINNFlow representation branches. Each representation is evaluated with a fixed downstream classifier and compared using clean-test metrics, confidence intervals, and paired significance tests.}
\label{fig:overall_architecture}
\end{figure*}
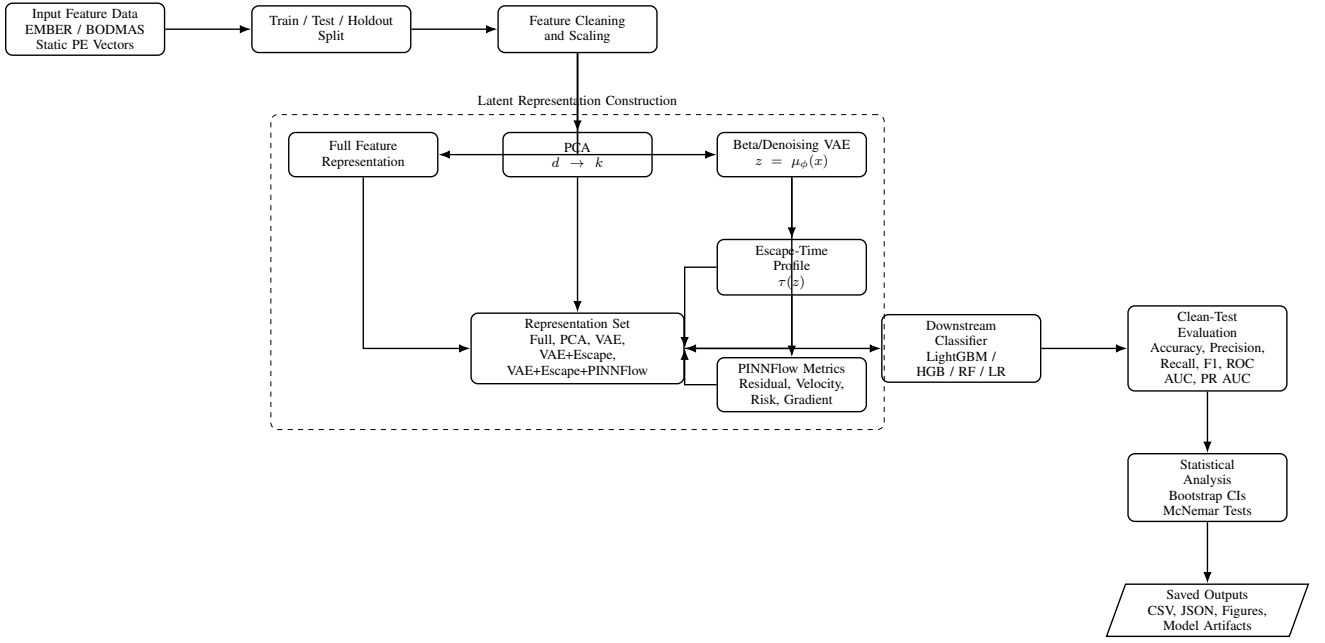

\begin{figure*}[!t]
\centering
\resizebox{0.96\textwidth}{!}{
\begin{tikzpicture}[
    >=Latex,
    node distance=1.35cm and 1.8cm,
    block/.style={
        rectangle,
        rounded corners,
        draw=black,
        thick,
        align=center,
        text width=3.5cm,
        minimum height=1.0cm,
        font=\footnotesize,
        fill=white
    },
    smallblock/.style={
        rectangle,
        rounded corners,
        draw=black,
        thick,
        align=center,
        text width=3.0cm,
        minimum height=0.95cm,
        font=\footnotesize,
        fill=white
    },
    output/.style={
        trapezium,
        trapezium left angle=70,
        trapezium right angle=110,
        draw=black,
        thick,
        align=center,
        text width=3.6cm,
        minimum height=0.95cm,
        font=\footnotesize,
        fill=white
    },
    arrow/.style={->, thick}
]

\node[block] (holdout) {Holdout Samples\\
$X_{\mathrm{hold}}, y_{\mathrm{hold}}$};

\node[block, right=1.6cm of holdout] (probe) {Perturbation Suite\\
Gaussian, Dropout,\\
Injection, Benign-Centroid,\\
Histogram Proxy};

\node[block, below=1.55cm of probe] (external) {Optional External PE Pairs\\
$X_{\mathrm{clean}}, X_{\mathrm{perturbed}}, y$};

\node[smallblock, right=1.7cm of probe] (encodeclean) {Encode Clean\\
Representations\\
$Z$};

\node[smallblock, right=1.7cm of external] (encodepert) {Encode Perturbed\\
Representations\\
$Z'$};

\node[block, right=2.0cm of encodeclean] (geom) {Geometric Stability\\
L2 displacement\\
Cosine similarity\\
kNN overlap};

\node[block, below=1.45cm of geom] (led) {Latent Escape Divergence\\
$\mathrm{LED}(x,x')=\frac{1}{K}\|\tau(z')-\tau(z)\|_1$};

\node[block, below=1.45cm of led] (pinn) {PINNFlow Shifts\\
Residual shift\\
Velocity shift\\
Risk-probability shift\\
Gradient shift};

\node[block, right=2.0cm of led] (robust) {Classifier Robustness\\
Predictions on $Z'$\\
F1, ROC AUC, PR AUC};

\node[output, below=1.5cm of robust] (tables) {Robustness and Stability\\
Tables / Figures};

\begin{scope}[on background layer]
\node[
    draw=black,
    dashed,
    rounded corners,
    fit=(probe)(external)(encodeclean)(encodepert),
    inner sep=0.35cm,
    label={[font=\footnotesize]above:Perturbation Input Layer}
] (inputgroup) {};

\node[
    draw=black,
    dashed,
    rounded corners,
    fit=(geom)(led)(pinn)(robust)(tables),
    inner sep=0.35cm,
    label={[font=\footnotesize]above:Stability and Robustness Evaluation}
] (evalgroup) {};
\end{scope}

\draw[arrow] (holdout) -- (probe);
\draw[arrow] (probe) -- (encodeclean);
\draw[arrow] (probe.east) |- (encodepert.west);
\draw[arrow] (external) -- (encodepert);

\draw[arrow] (probe) -- (external);

\draw[arrow] (encodeclean) -- (geom);
\draw[arrow] (encodepert.north) |- (geom.south west);

\draw[arrow] (encodeclean.east) |- (led.west);
\draw[arrow] (encodepert) -- (led);

\draw[arrow] (encodepert.east) |- (pinn.west);

\draw[arrow] (encodepert.east) -| (robust.west);

\draw[arrow] (geom.east) -| (tables.north);
\draw[arrow] (led.east) -- (robust.west);
\draw[arrow] (pinn.east) -| (tables.west);
\draw[arrow] (robust) -- (tables);

\end{tikzpicture}
}
\caption{Perturbation-stability and external PE-pair evaluation workflow. Holdout samples are perturbed using controlled feature-space probes, while externally generated PE-level clean/perturbed feature pairs may also be supplied. Clean and perturbed representations are compared using geometric stability metrics, Latent Escape Divergence, PINNFlow shift metrics, and downstream classifier robustness.}
\label{fig:perturbation_evaluation}
\end{figure*}
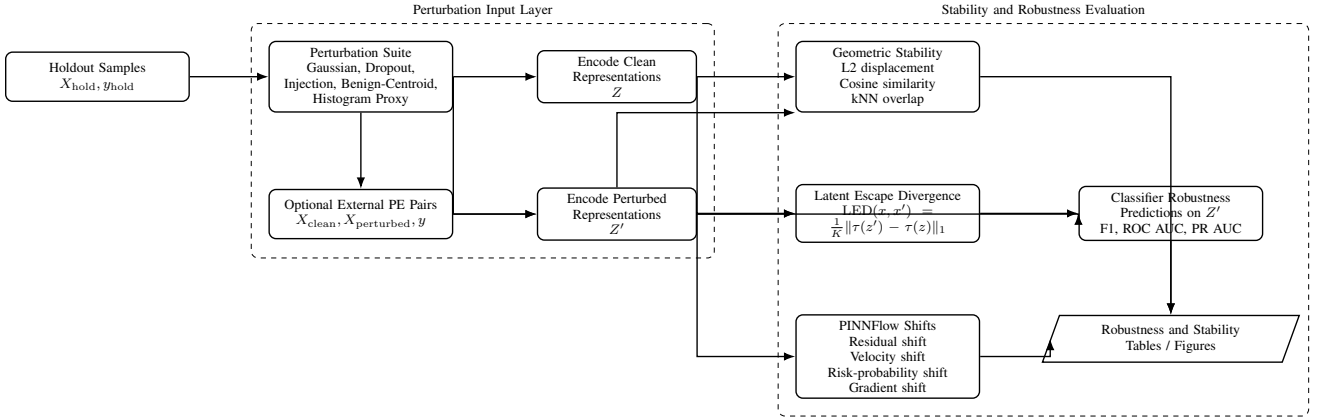

\section{Results}
\subsection{VAE and PINNFlow Training Diagnostics}
\begin{figure*}[t]
\centering
\subfloat[VAE training curve.]{\includegraphics[width=0.48\textwidth]{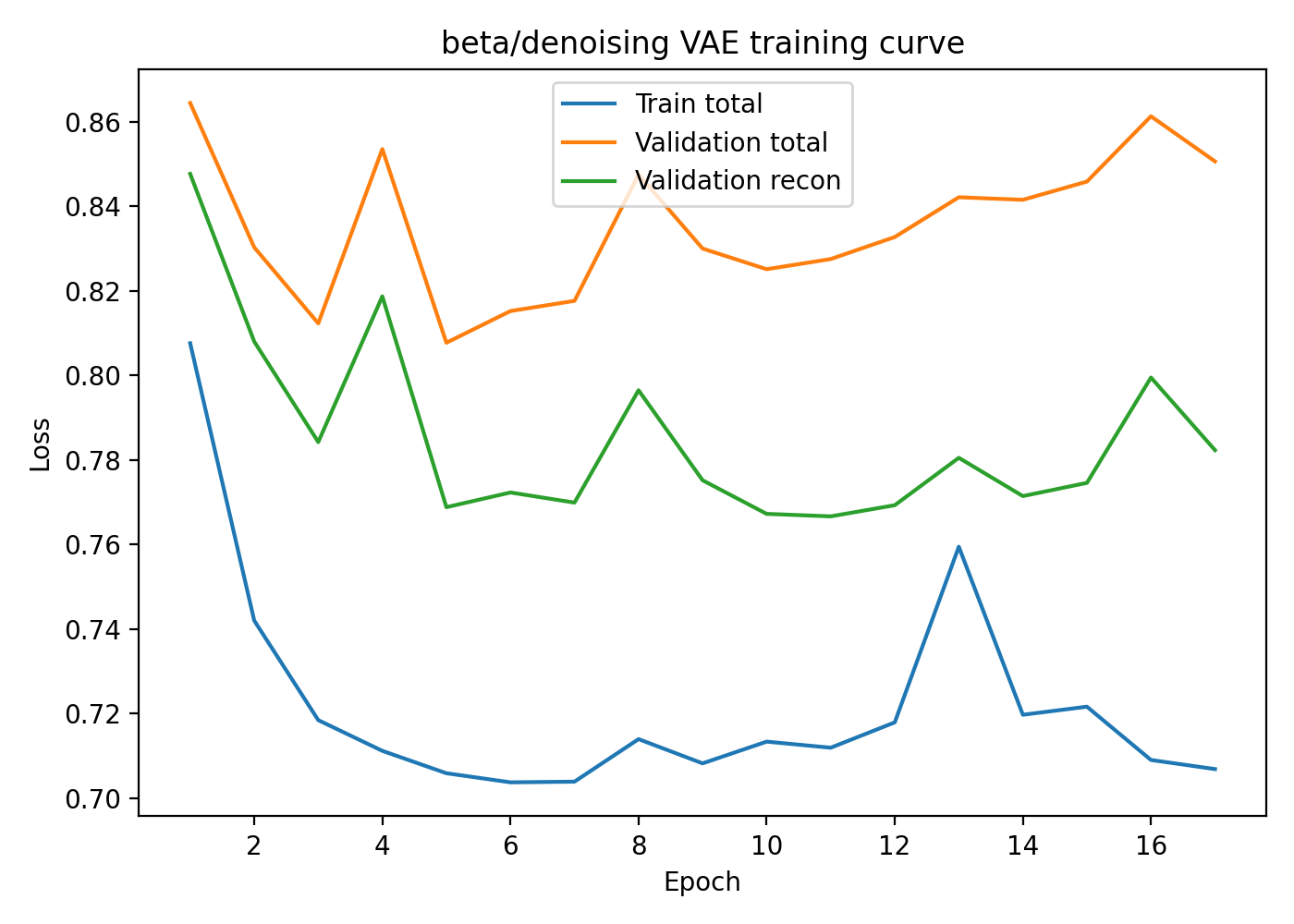}}
\hfill
\subfloat[PINNFlow training diagnostics.]{\includegraphics[width=0.48\textwidth]{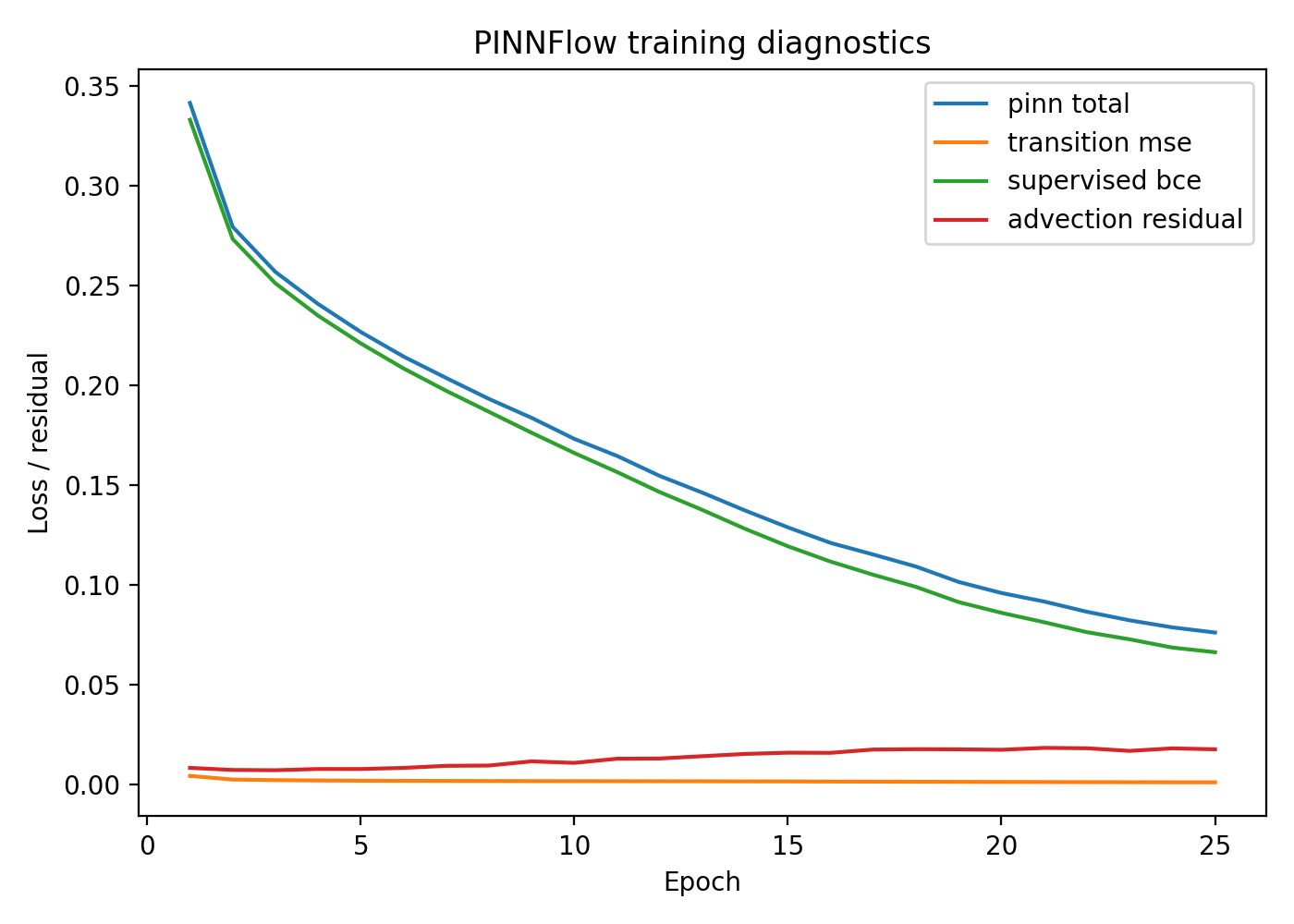}}
\caption{Training diagnostics. The VAE avoids the earlier near-zero-KL collapse, and PINNFlow learns transition, supervised, residual, and energy terms.}
\label{fig:training}
\end{figure*}

The VAE configuration directly addresses the earlier KL-collapse weakness. The final latent diagnostics report mean total KL per sample of 14.83, mean KL per dimension of 0.232, 64 active units using variance $>10^{-3}$, 55 active units using variance $>10^{-2}$, and no KL-collapse flag. Fig.~\ref{fig:training} shows the VAE training curve and PINNFlow diagnostics. PINNFlow total loss decreases from 0.341 to 0.076, supervised BCE decreases from 0.333 to 0.066, and transition MSE decreases from 0.00427 to 0.00105. This does not prove superior classification, but it shows that PINNFlow learned measurable latent-trajectory structure.

\begin{table}[t]
\centering
\caption{VAE latent diagnostics }
\label{tab:vae_diag}
\footnotesize
\begin{tabular}{lr}
\toprule
Diagnostic & Value \\
\midrule
Latent dimension & 64 \\
Mean total KL per sample & 14.830 \\
Mean KL per dimension & 0.232 \\
Median KL per dimension & 0.0076 \\
Active units, var $>10^{-3}$ & 64 \\
Active units, var $>10^{-2}$ & 55 \\
Mean latent mean variance & 0.2015 \\
KL collapse flag & False \\
\bottomrule
\end{tabular}
\end{table}

\subsection{Clean Classification Ablation}
\begin{figure*}[t]
\centering
\subfloat[ROC ablation.]{\includegraphics[width=0.48\textwidth]{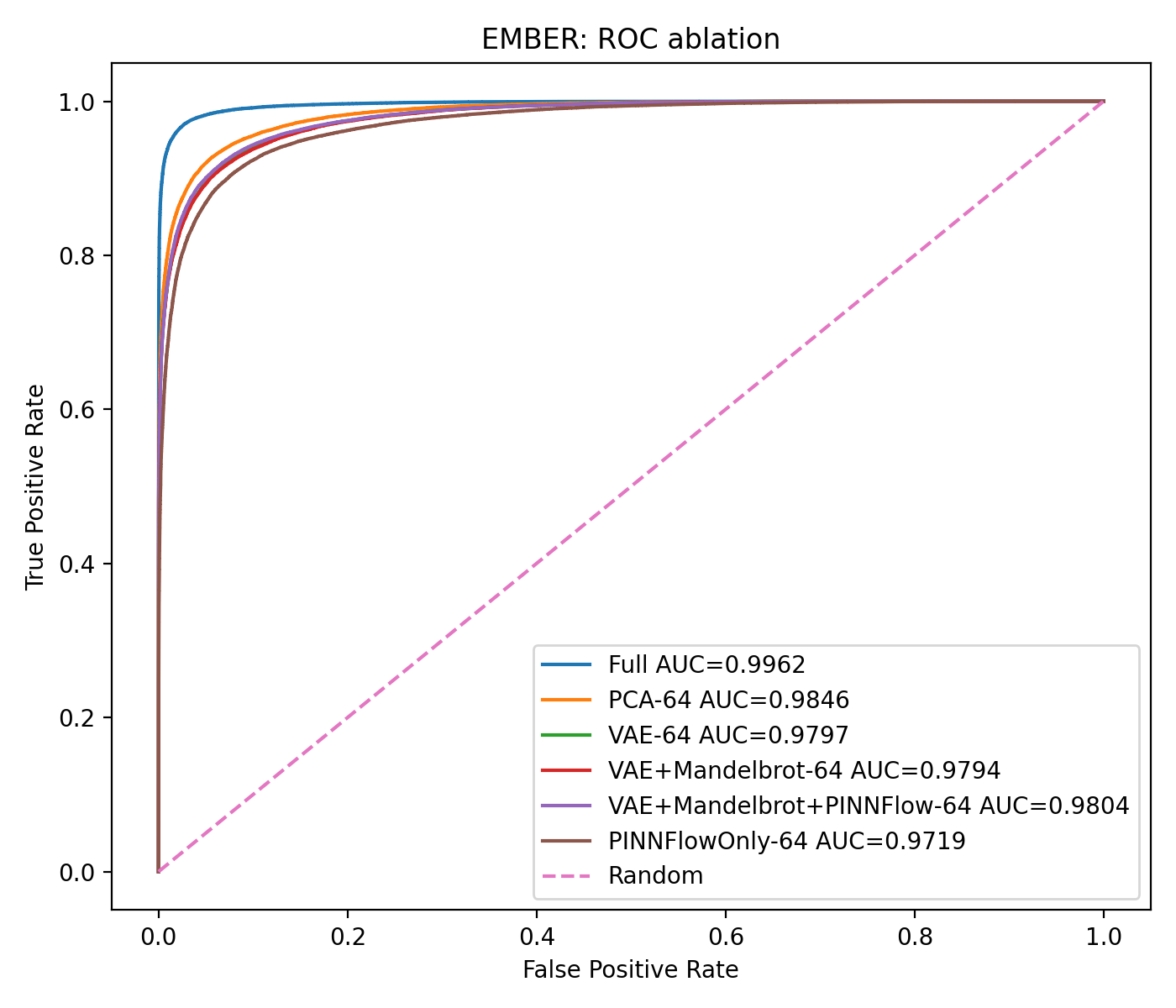}}
\hfill
\subfloat[Precision-recall ablation.]{\includegraphics[width=0.48\textwidth]{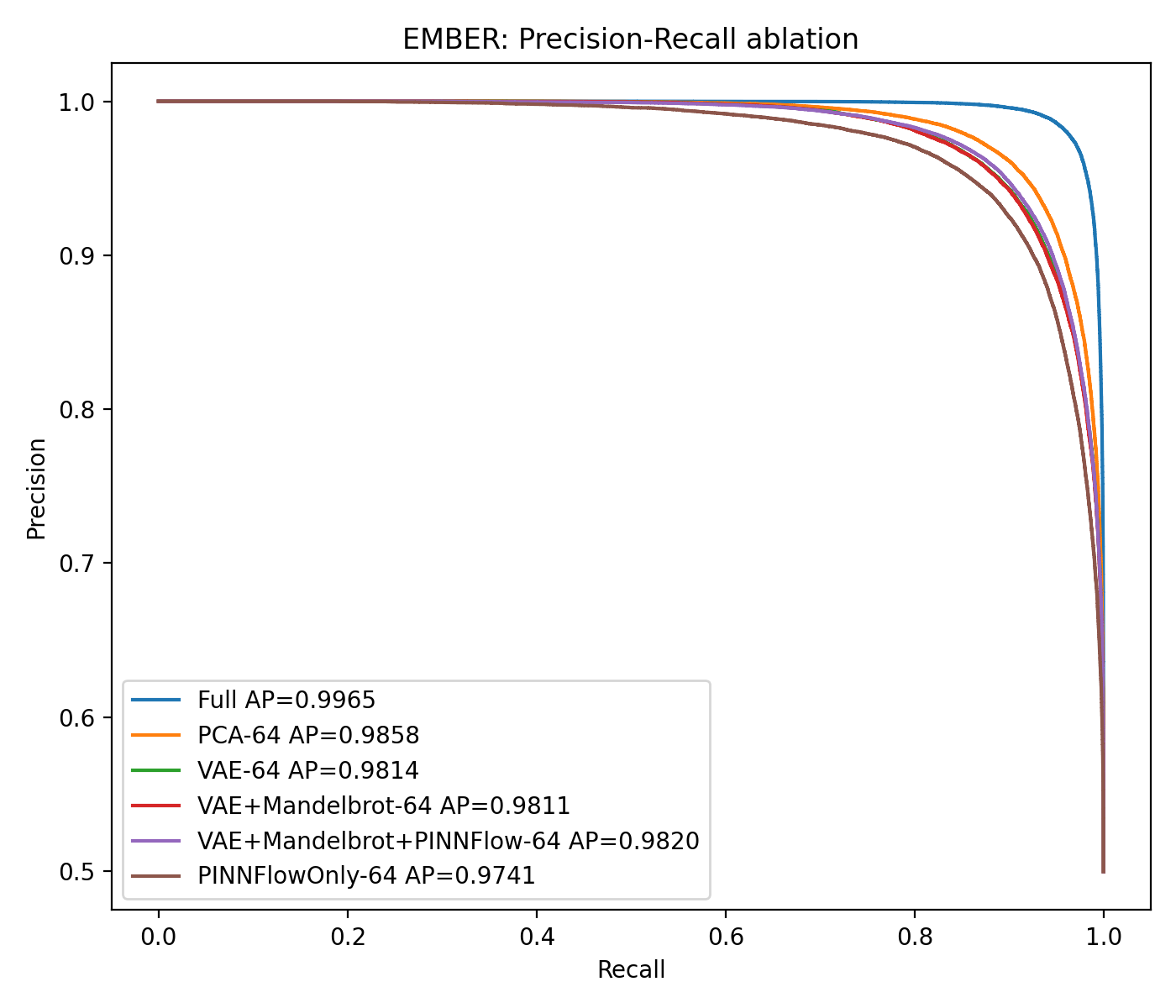}}
\caption{Clean classification ablation. Full features and PCA-64 remain strongest, while VAE+Mandelbrot+PINNFlow-64 is best among the VAE-derived variants.}
\label{fig:aucpr}
\end{figure*}

Table~\ref{tab:clean} reports clean EMBER classification. Full features remain strongest. PCA-64 is the strongest compressed representation. The VAE, Mandelbrot, and PINNFlow variants improve the diagnostic expressiveness of the latent space but do not surpass PCA-64 for clean classification. Fig.~\ref{fig:aucpr} shows the ROC and precision-recall ablations.

\begin{table*}[t]
\centering
\caption{Clean EMBER classification ablation.}
\label{tab:clean}
\scriptsize
\begin{tabular}{lrrrrrrrr}
\toprule
Method & Features & Accuracy & Precision & Recall & F1 & ROC AUC & PR AUC & Train s \\
\midrule
Full & 2381 & 0.9714 & 0.9741 & 0.9685 & 0.9713 & 0.9962 & 0.9965 & 372.60 \\
PCA-64 & 64 & 0.9350 & 0.9397 & 0.9297 & 0.9347 & 0.9846 & 0.9858 & 19.54 \\
VAE+Mandelbrot+PINNFlow-64 & 113 & 0.9262 & 0.9313 & 0.9202 & 0.9257 & 0.9804 & 0.9820 & 22.77 \\
VAE-64 & 64 & 0.9237 & 0.9270 & 0.9199 & 0.9234 & 0.9797 & 0.9814 & 15.95 \\
VAE+Mandelbrot-64 & 107 & 0.9227 & 0.9258 & 0.9189 & 0.9224 & 0.9794 & 0.9811 & 26.10 \\
PINNFlowOnly-64 & 6 & 0.9137 & 0.9179 & 0.9086 & 0.9133 & 0.9719 & 0.9741 & 3.72 \\
\bottomrule
\end{tabular}
\end{table*}

\subsection{Latent Geometry and Escape-Time Behavior}
\begin{figure*}[t]
\centering
\subfloat[PCA-64 projection.]{\includegraphics[width=0.48\textwidth]{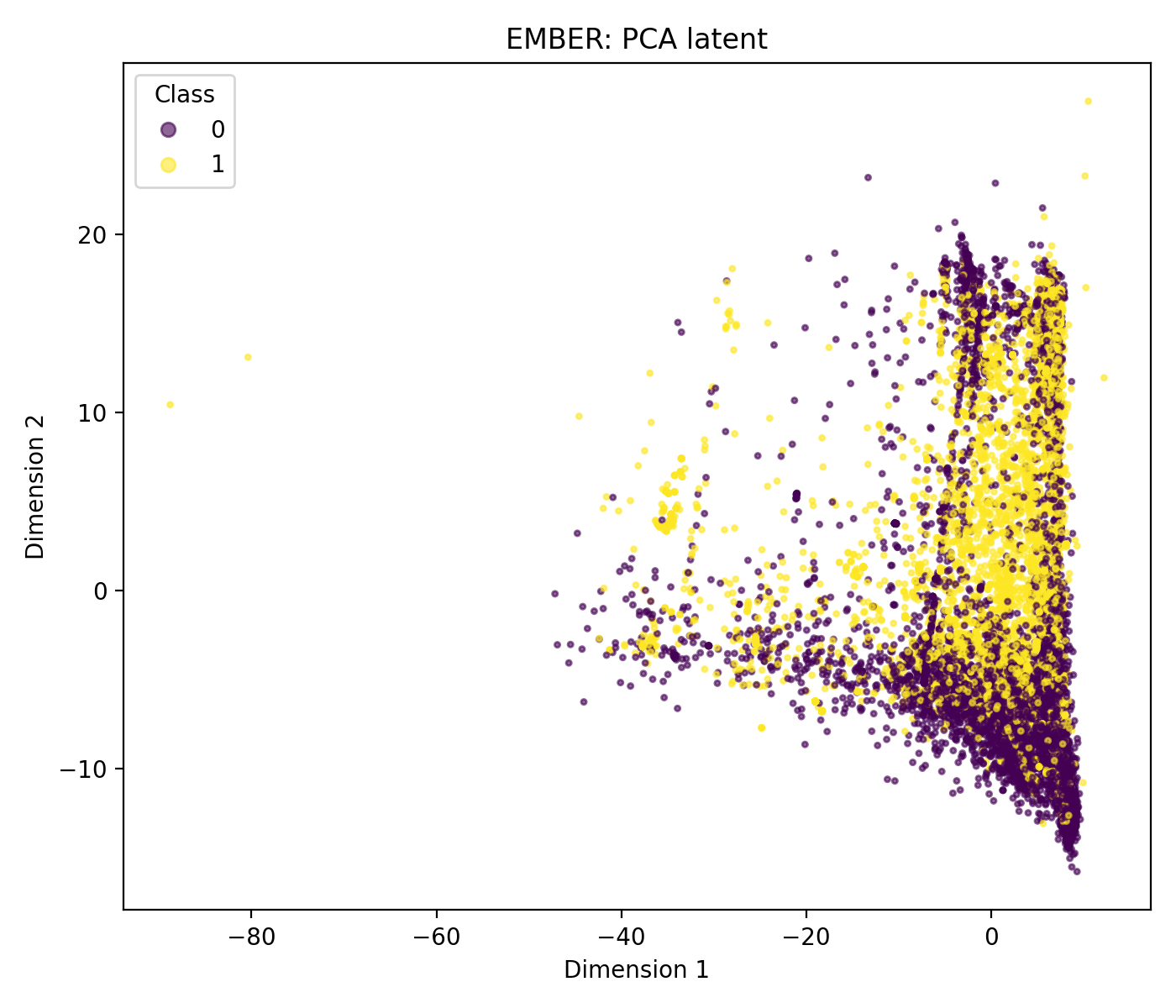}}
\hfill
\subfloat[VAE-64 projection.]{\includegraphics[width=0.48\textwidth]{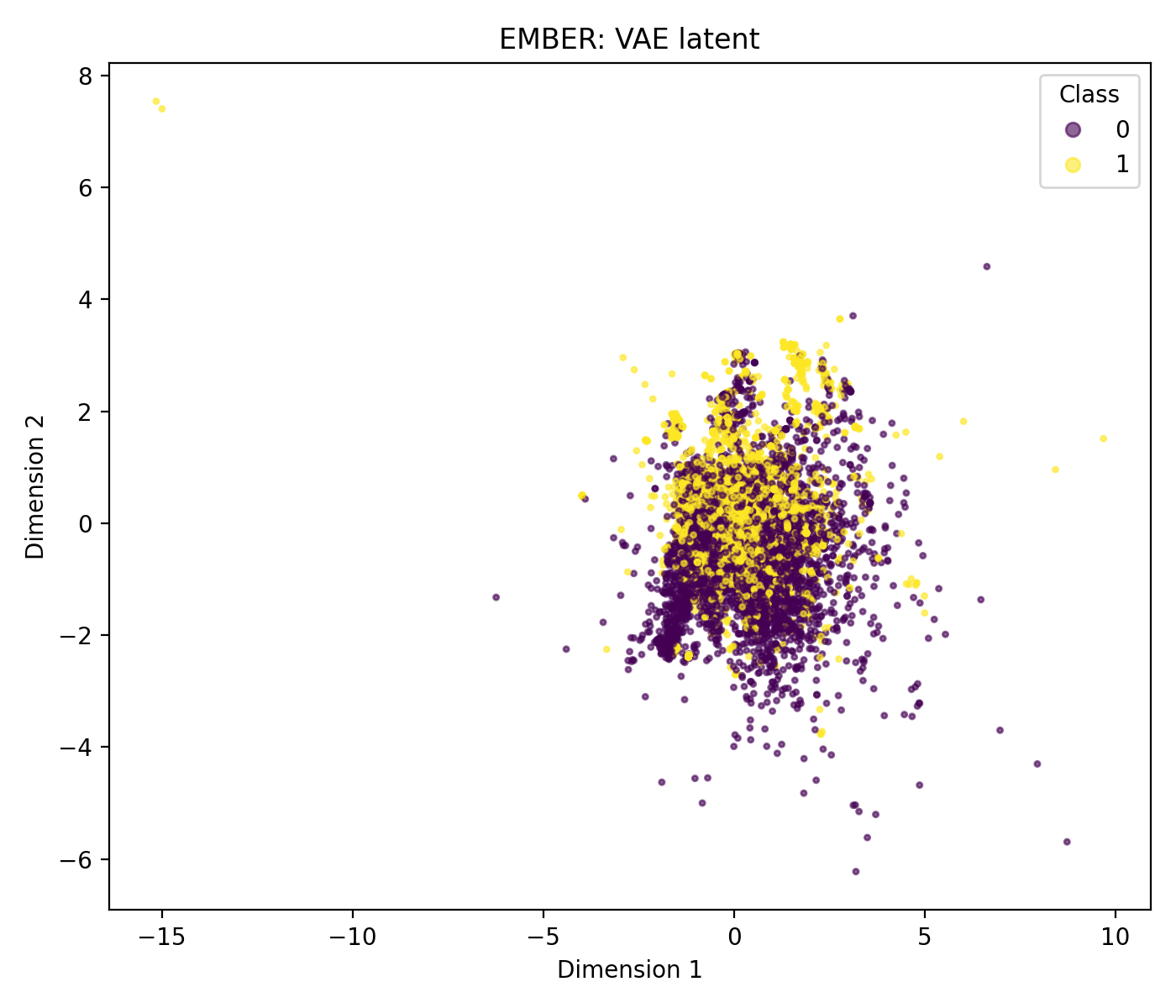}}
\caption{Latent visualizations for PCA and VAE representations.}
\label{fig:latents1}
\end{figure*}

\begin{figure*}[t]
\centering
\subfloat[VAE+Mandelbrot-64 projection.]{\includegraphics[width=0.48\textwidth]{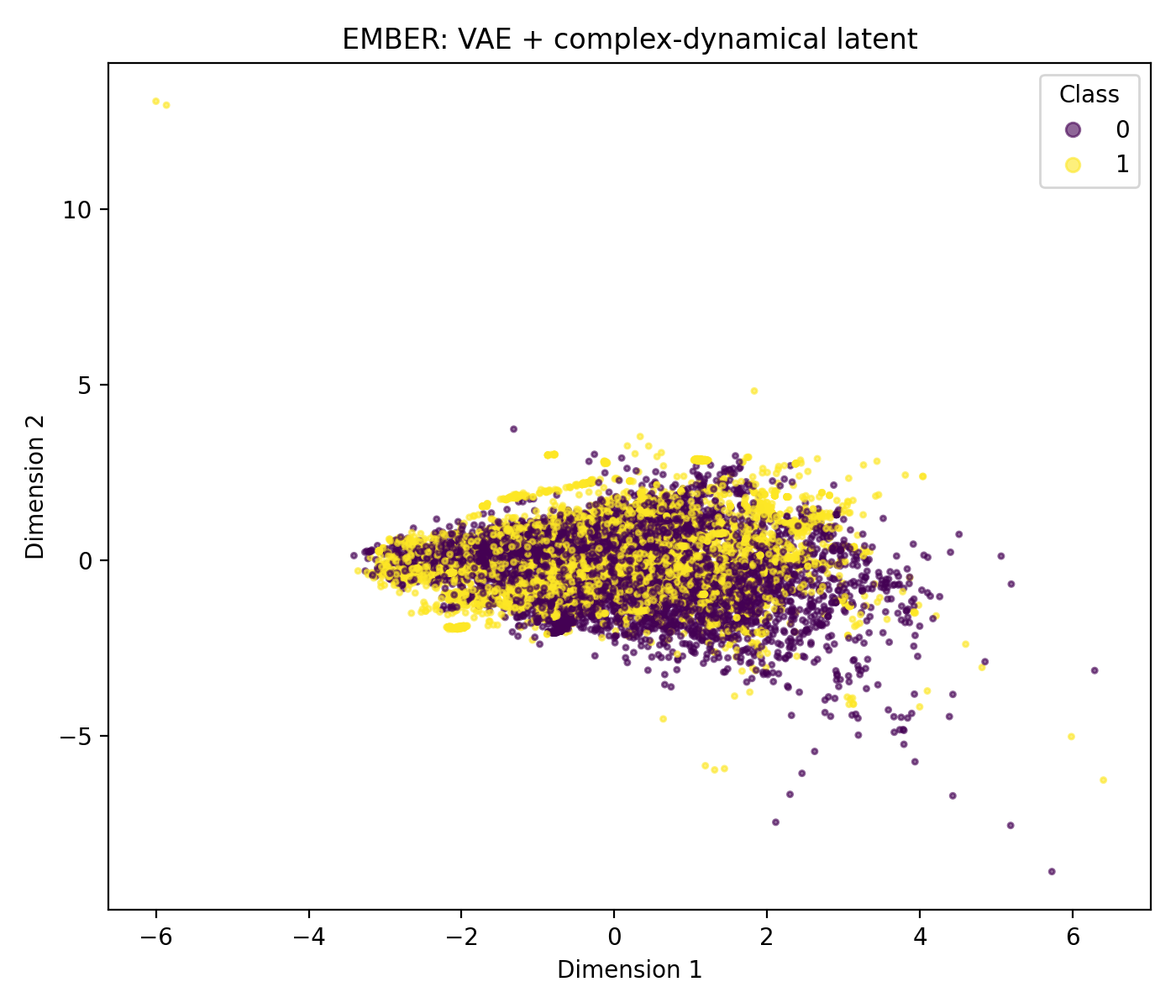}}
\hfill
\subfloat[VAE+Mandelbrot+PINNFlow-64 projection.]{\includegraphics[width=0.48\textwidth]{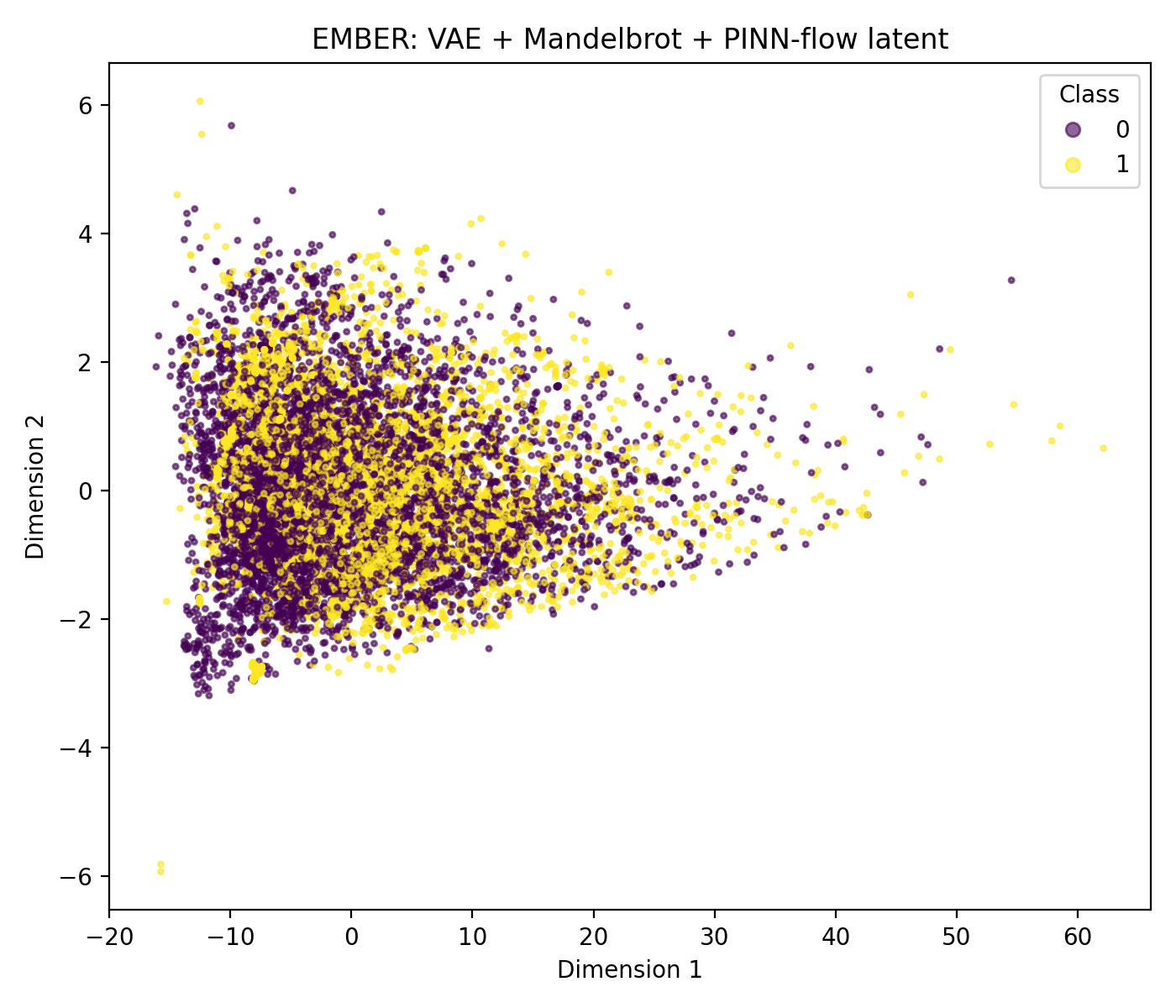}}
\caption{Complex-dynamical and PINNFlow latent visualizations. PINNFlow changes the geometry but does not eliminate class overlap.}
\label{fig:latents2}
\end{figure*}

\begin{figure}[t]
\centering
\includegraphics[width=\columnwidth]{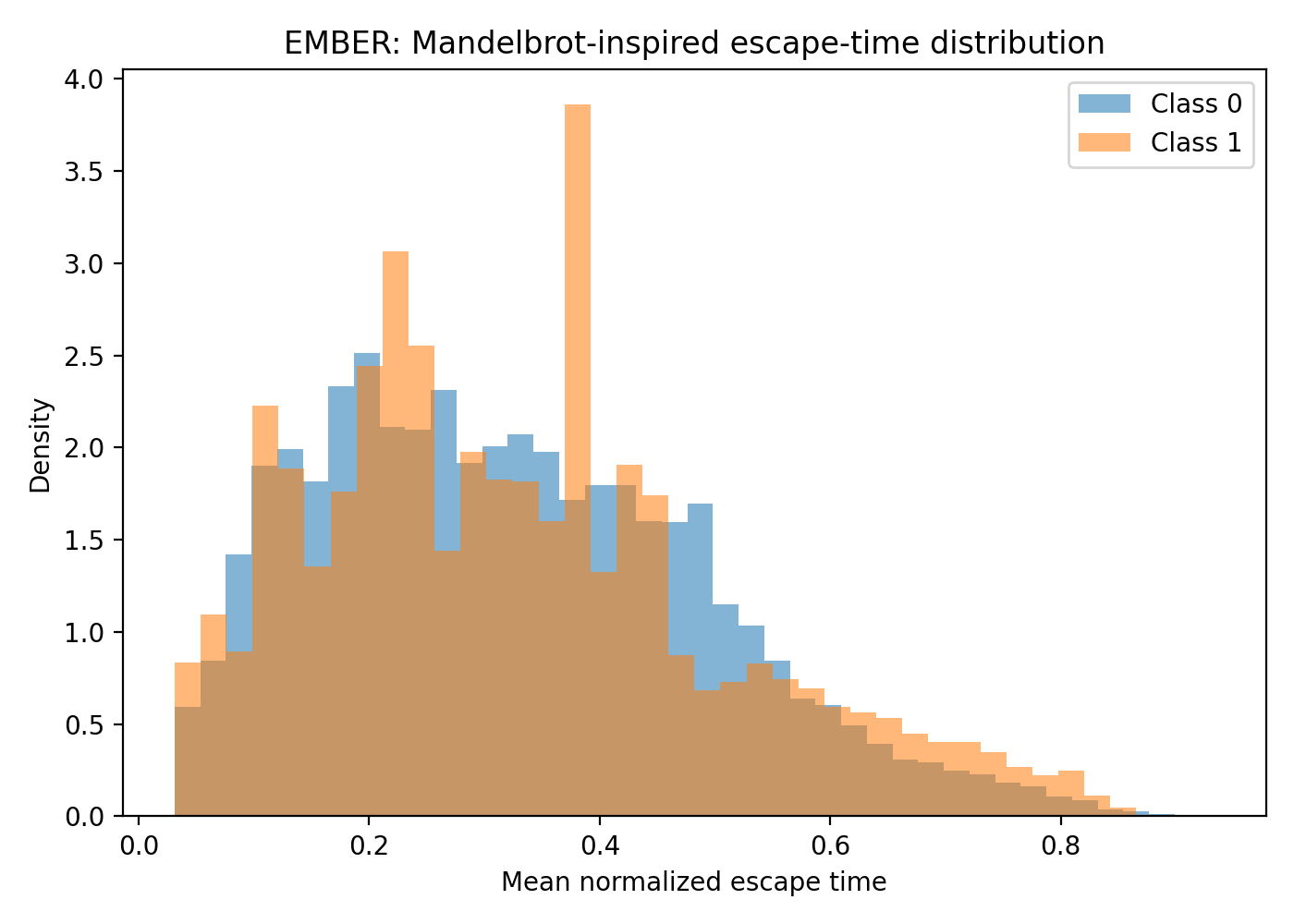}
\caption{Distribution of mean normalized Mandelbrot-inspired escape time by class. Escape profiles overlap, supporting their use as stability descriptors rather than direct labels.}
\label{fig:escape}
\end{figure}

Figs.~\ref{fig:latents1} and \ref{fig:latents2} show two-dimensional projections of the learned representations. PCA preserves stronger class structure than the VAE latent. The VAE+Mandelbrot and VAE+Mandelbrot+PINNFlow projections reshape the geometry and expose additional structure, but class overlap remains substantial. Fig.~\ref{fig:escape} shows the distribution of mean normalized escape time by class; the overlap confirms that escape time is not a standalone classifier, but a stability descriptor.

\subsection{Boundary Complexity}
Table~\ref{tab:boundary} reports box-counting complexity for uncertain decision regions. PINNFlowOnly has the largest estimated boundary dimension, followed by PCA-64 and VAE+Mandelbrot+PINNFlow-64. The VAE+Mandelbrot representation has higher boundary complexity than VAE-64, supporting the claim that complex-dynamical features expose additional nonlinear structure in uncertain regions. The high $R^2$ values indicate stable log-log scaling fits for most representations.

\begin{table}[t]
\centering
\caption{Boundary box-counting complexity on uncertain decision regions.}
\label{tab:boundary}
\footnotesize
\begin{tabular}{lrr}
\toprule
Method & Dimension & $R^2$ \\
\midrule
Full & 1.6280 & 0.9947 \\
PCA-64 & 1.7164 & 0.9952 \\
VAE+Mandelbrot+PINNFlow-64 & 1.7084 & 0.9958 \\
VAE-64 & 1.4413 & 0.9934 \\
VAE+Mandelbrot-64 & 1.5517 & 0.9949 \\
PINNFlowOnly-64 & 1.7902 & 0.9960 \\
\bottomrule
\end{tabular}
\end{table}

\subsection{Perturbation Robustness}
Table~\ref{tab:robust} reports F1 under perturbation probes. PCA-64 remains strongest overall. However, VAE+Mandelbrot+PINNFlow-64 improves over VAE+Mandelbrot-64 for all five probes and over VAE-64 for several structured probes. The gain is largest for movement toward the benign centroid (0.9057 to 0.9145) and EMBER histogram proxy perturbation (0.9105 to 0.9159). This is the strongest empirical support for PINNFlow: it does not improve clean classification over PCA, but it partially recovers robustness relative to the complex-dynamical VAE under structured perturbations.

\begin{table*}[t]
\centering
\caption{Perturbation robustness measured by F1.}
\label{tab:robust}
\scriptsize
\begin{tabular}{lrrrrr}
\toprule
Perturbation & PCA-64 & VAE-64 & VAE+Mandelbrot-64 & VAE+Mandelbrot+PINNFlow-64 & PINNFlowOnly-64 \\
\midrule
Gaussian small & 0.9318 & 0.9225 & 0.9228 & 0.9245 & 0.9104 \\
Sparse dropout & 0.9275 & 0.9163 & 0.9170 & 0.9205 & 0.9085 \\
Sparse injection & 0.9331 & 0.9230 & 0.9243 & 0.9281 & 0.9114 \\
Toward benign centroid & 0.9229 & 0.9072 & 0.9057 & 0.9145 & 0.9053 \\
EMBER histogram proxy & 0.9223 & 0.9091 & 0.9105 & 0.9159 & 0.9021 \\
\bottomrule
\end{tabular}
\end{table*}

\subsection{LED and PINN-Specific Perturbation Diagnostics}
\begin{figure}[t]
\centering
\includegraphics[width=\columnwidth]{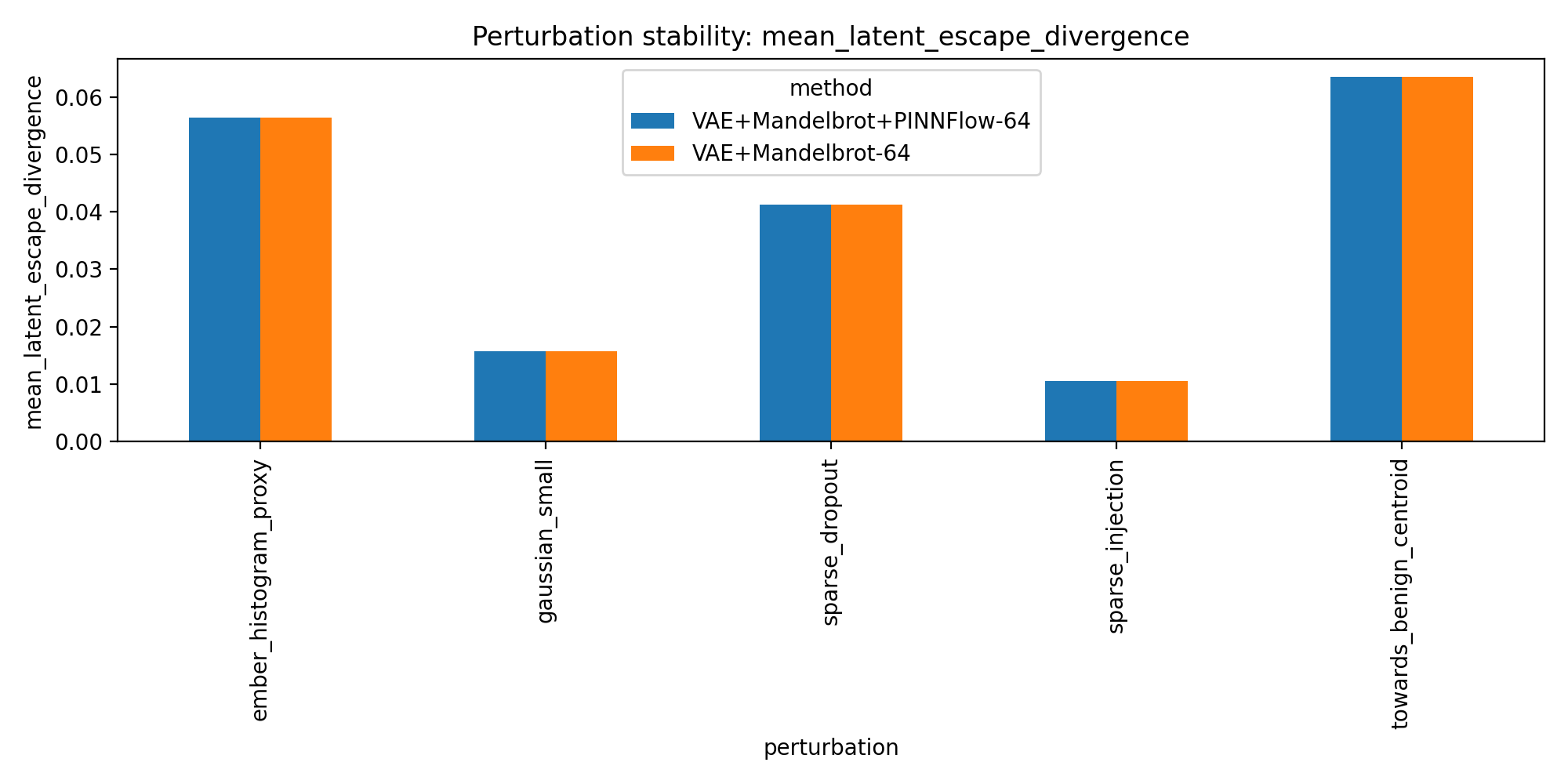}
\caption{Mean Latent Escape Divergence under perturbation probes. LED captures perturbation-induced change in Mandelbrot escape-time fingerprints.}
\label{fig:led}
\end{figure}

Fig.~\ref{fig:led} and Table~\ref{tab:led} report LED. The structured perturbations that most change the escape profile are benign-centroid movement and the EMBER histogram proxy, followed by sparse dropout. Sparse injection and small Gaussian noise produce smaller escape-time shifts. Because LED is computed from the Mandelbrot escape profile, it is identical for VAE+Mandelbrot and VAE+Mandelbrot+PINNFlow; this is expected and motivates separate PINNFlow-specific diagnostics.

\begin{table}[t]
\centering
\caption{Mean LED by perturbation.}
\label{tab:led}
\footnotesize
\begin{tabular}{lr}
\toprule
Perturbation & Mean LED \\
\midrule
Gaussian small & 0.0157 \\
Sparse dropout & 0.0412 \\
Sparse injection & 0.0105 \\
Toward benign centroid & 0.0635 \\
EMBER histogram proxy & 0.0564 \\
\bottomrule
\end{tabular}
\end{table}

\begin{figure*}[t]
\centering
\subfloat[PINN residual shift.]{\includegraphics[width=0.32\textwidth]{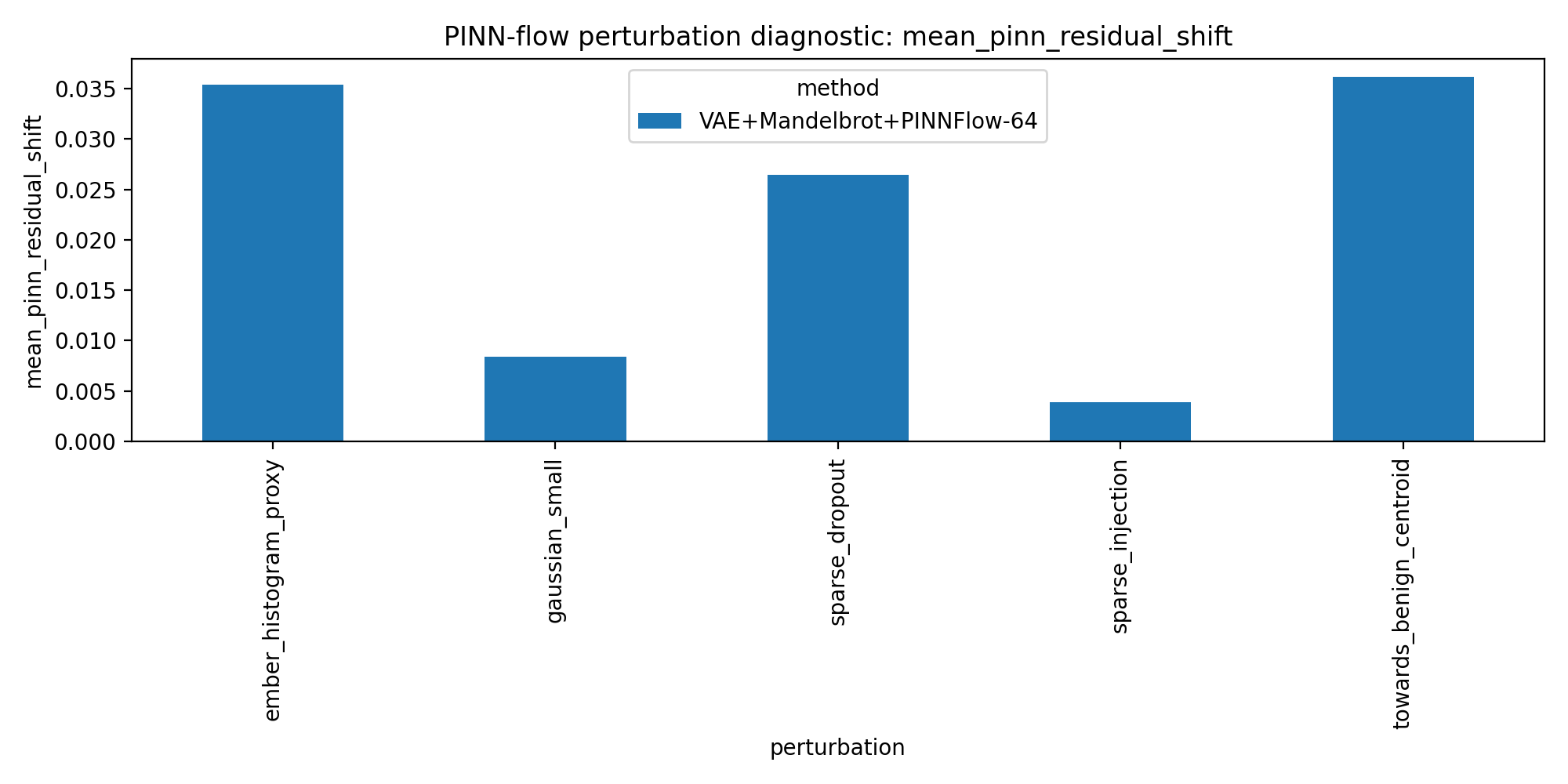}}
\hfill
\subfloat[PINN risk-probability shift.]{\includegraphics[width=0.32\textwidth]{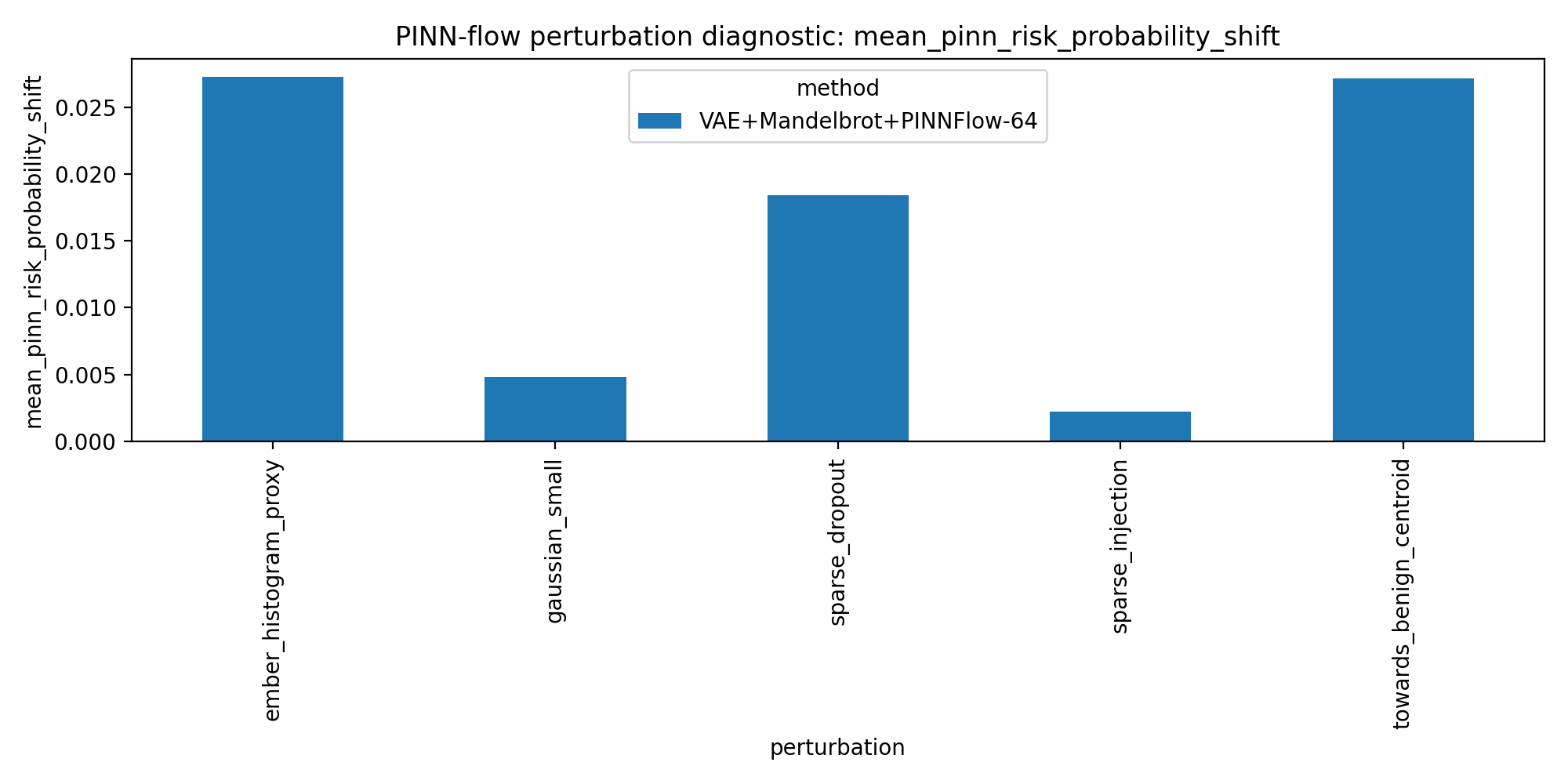}}
\hfill
\subfloat[PINN velocity-norm shift.]{\includegraphics[width=0.32\textwidth]{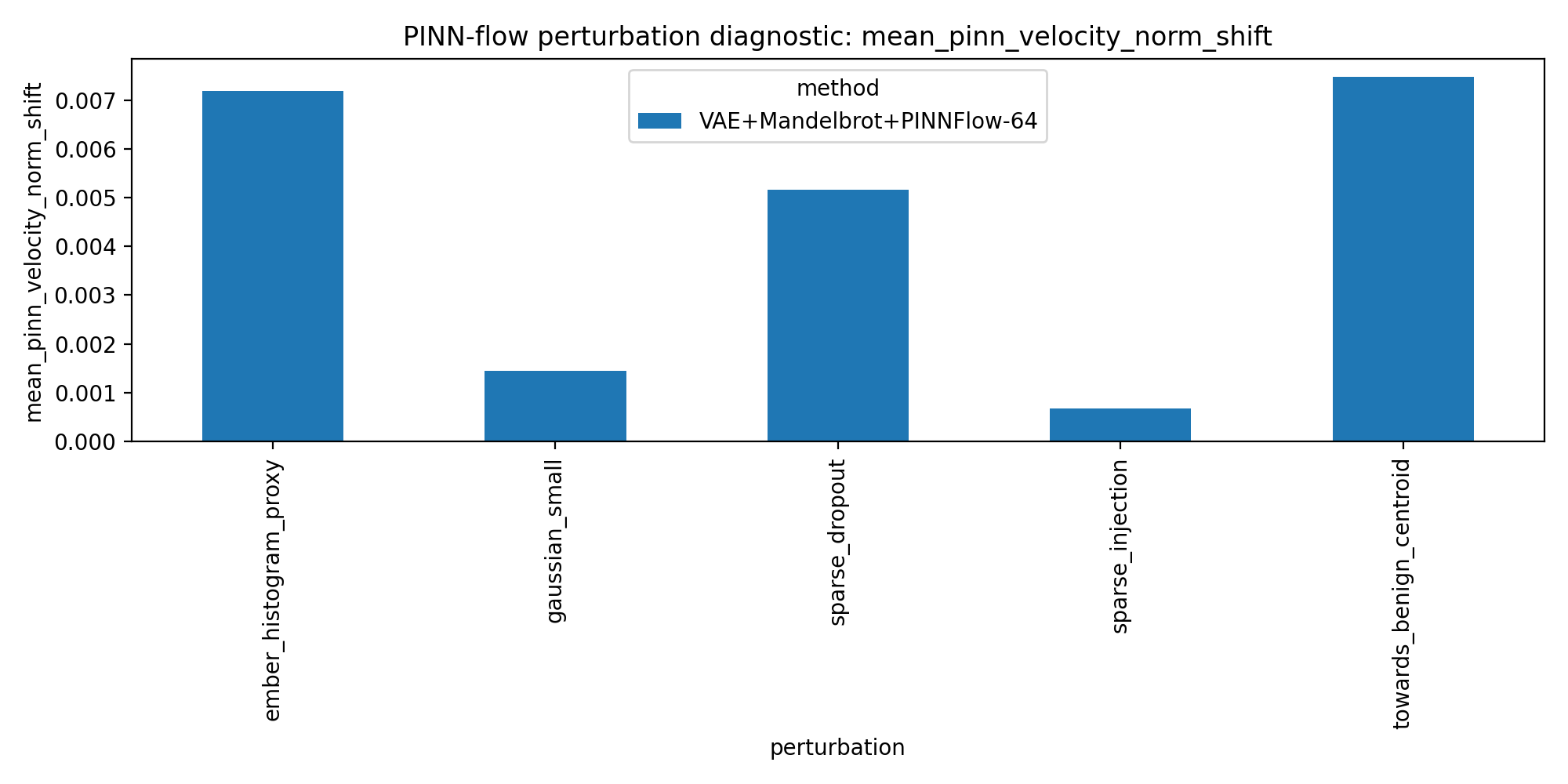}}
\caption{PINNFlow-specific perturbation diagnostics. These features isolate the PINN contribution, unlike LED which is driven by the Mandelbrot escape profile.}
\label{fig:pinnshifts}
\end{figure*}

The code adds PINN-specific metrics. Fig.~\ref{fig:pinnshifts} shows residual, risk-probability, and velocity-norm shifts. These plots make the PINNFlow contribution measurable beyond LED. The benign-centroid and EMBER histogram perturbations produce the largest PINN residual and risk shifts, aligning with the perturbation probes that resemble evasion-like movement toward benign-looking regions.

\begin{table}[H]
\centering
\caption{Selected PINNFlow perturbation shifts for VAE+Mandelbrot+PINNFlow-64.}
\label{tab:pinnshifts}
\footnotesize
\begin{tabular}{lrrr}
\toprule
Perturbation & Residual & Risk & Velocity \\
\midrule
Gaussian small & 0.0084 & 0.0048 & 0.0014 \\
Sparse dropout & 0.0264 & 0.0184 & 0.0052 \\
Sparse injection & 0.0039 & 0.0022 & 0.0007 \\
Toward benign centroid & 0.0362 & 0.0272 & 0.0075 \\
EMBER histogram proxy & 0.0354 & 0.0273 & 0.0072 \\
\bottomrule
\end{tabular}
\end{table}

\subsection{Paired Significance Tests}
McNemar tests compare VAE-64 with other methods. Table~\ref{tab:mcnemar} shows significant differences for all comparisons. The direction matters: Full and PCA-64 are significantly stronger than VAE-64, while VAE+Mandelbrot+PINNFlow-64 is significantly different and improves over VAE-64 in clean metrics but does not surpass PCA-64 or Full features.

\begin{table}[H]
\centering
\caption{McNemar paired tests using VAE-64 as baseline.}
\label{tab:mcnemar}
\footnotesize
\begin{tabular}{lrrr}
\toprule
Comparison & $b$ & $c$ & $p$ \\
\midrule
VAE-64 vs Full & 1630 & 10209 & $<10^{-6}$ \\
VAE-64 vs PCA-64 & 3247 & 5283 & $<10^{-6}$ \\
VAE-64 vs VAE+Mandelbrot-64 & 1774 & 1587 & 0.0013 \\
VAE-64 vs VAE+Mandelbrot+PINNFlow-64 & 4748 & 5193 & $<10^{-5}$ \\
VAE-64 vs PINNFlowOnly-64 & 7179 & 5381 & $<10^{-6}$ \\
\bottomrule
\end{tabular}
\end{table}

\section{Discussion}
\subsection{Interpreting the Latent-Dynamics Contribution}
The results support four claims. First, latent dynamics matter because perturbations induce paths in representation space, and those paths expose stability properties invisible to clean metrics. Second, LED provides a mathematically defined stability metric: the average absolute change in normalized complex-dynamical escape profile. Third, PINNFlow enforces structure through a transport-style residual and yields measurable residual/velocity/risk-shift diagnostics. Fourth, fractal descriptors capture boundary complexity by estimating how uncertain decision regions scale across box sizes.

\subsection{Boundaries of the Claims}
The method should not be presented as a superior clean classifier. Full features remain strongest; PCA-64 remains the best compressed baseline. It should also not be claimed that malware is governed by Mandelbrot dynamics or physical laws. The correct framing is that complex dynamics and PINN-style residuals are mathematical descriptors and inductive biases for representation stability analysis.

\subsection{Methodological Safeguards and Threats to Validity}
The pipeline directly addresses four expected concerns. VAE undertraining was addressed through KL warm-up, free bits, lower beta, larger latent dimension, and active-unit diagnostics. PINNFlow's unique contribution was made measurable through PINNFlowOnly, residual shift, velocity shift, risk shift, and gradient-shift diagnostics. Functionality preservation is explicitly treated as a limitation, with an external perturbation-pair interface for future PE-level workflows. The Mandelbrot component is mathematically grounded as escape-time sensitivity analysis and boundary-complexity characterization, not as metaphor.

\section{Generalization Beyond Malware Detection}

Although this study focuses on static malware detection, the proposed latent-stability perspective may be relevant to other security-critical machine learning tasks where structured input changes can alter model confidence or representation geometry. Examples include intrusion detection, phishing classification, fraud analytics, and anomaly-based network monitoring. However, this transfer should be treated as a hypothesis rather than an empirical conclusion of this paper.

The most portable components of the framework are the diagnostic ideas: measuring representation displacement, neighborhood stability, Latent Escape Divergence (LED), and latent-flow residual shifts under controlled perturbations. These diagnostics could be adapted to other domains if domain-appropriate perturbation operators are defined and validated. For example, a perturbation in phishing detection would need to preserve the semantic intent of a message, while a perturbation in intrusion detection would need to preserve the operational meaning of a network event.

Therefore, the contribution of this paper is not a general proof that LED or PINNFlow will improve robustness across all security domains. Rather, it provides a methodology that can be re-tested in other domains using task-specific perturbation models, datasets, and validation criteria.

\section{Limitations}
The most important limitation is that the default perturbations are feature-space probes. They are useful for controlled analysis but do not prove functionality-preserving PE manipulation. Second, the evaluation uses EMBER 2018-style static PE features; additional evaluation on BODMAS temporal data and EMBER2024 challenge samples is needed. Third, while the VAE does not collapse according to KL diagnostics, it still underperforms PCA-64 for clean classification. Fourth, PINNFlow improves robustness relative to VAE+Mandelbrot under several probes but still does not beat PCA-64. Fifth, box-counting dimensions are descriptive estimates and should be interpreted as boundary-complexity diagnostics rather than definitive fractal proofs.

\section{Future Work}
Future work should connect the pipeline to raw PE perturbation tools and feature re-extraction workflows, such as secml-malware and LIEF-style binary rewriting \cite{demetrio2021secml}. The strongest next experiment is to evaluate clean/perturbed pairs generated by functionality-preserving transformations and test whether LED and PINNFlow residual divergence detect evasion better than classifier confidence, PCA distance, or ordinary VAE latent distance. Additional work should include EMBER2024 challenge-set detection at fixed false-positive rates, BODMAS temporal transfer, contrastive latent learning, and calibration analysis for LED/PINN residual thresholds.

\section{Conclusion}
This paper presents a VAE+Mandelbrot+PINNFlow malware representation pipeline as a principled latent-dynamics framework. The  EMBER run uses anti-collapse VAE training, formal LED, PINNFlow transport residuals, PINN-specific diagnostics, and box-counting boundary analysis. Clean classification results show that Full features and PCA-64 remain superior. Nevertheless, the framework contributes a novel way to analyze malware perturbation behavior: it measures how representations move, how escape-time stability changes, how latent-flow residuals respond, and how uncertain decision boundaries scale. The contribution is therefore best understood as neural-geometric, complex-dynamical, and dynamics-informed perturbation-stability analysis for malware representation learning.

\section{Code Availability}

The source code used for the experiments in this study is available at:\url{https://github.com/abamidele/Latent-Stability-Analysis-of-Malware-Representations-Under-Feature-Space-Perturbations}.

\end{document}